# Microscope Projection Photolithography Based on Ultraviolet Light-emitting Diodes

Minjae Kwon[1] and Young-Gu Ju[2]

[1]Daegu Science High School, 154, Dongdaegu-ro, Suseong-gu, Daegu, 42110, Korea

[2]Department of Physics Education, Kyungpook National University, 80 Daehakro, Bukgu, Daegu, 41566, Korea

E-mail: ygju@knu.ac.kr

**Abstract**

We adapted a conventional microscope for projection photolithography using an ultraviolet (UV) light-emitting diode (LED) as a light source. The use of a UV LED provides the microscope projector with several advantages in terms of compactness and cost. The adapted microscope was capable of producing line patterns as wide as 5 μm with the use of a 4X objective lens under optimal lithography conditions. The obtained line width is close to that of the diffraction limit, implying that the line width can be reduced further with the use of a higher resolution photomask and higher magnification objective lens. We expect that low-cost microscope projection photolithography based on a UV LED contribute to the field of physics education or various areas of research, such as chemistry and biology, in the future.

## 1. Introduction

Photolithography, or optical lithography, is one of the key techniques used in the fabrication of an integrated circuit[1]. Recently, the applications of photolithography have spread to various research fields, such as microelectromechanical systems (MEMS), micro-fluidics, bio-sensors, biology, chemistry, physics, and material science, with an interest in generating small areas of patterned microstructures. Therefore, many academic institutes are equipped with mask-aligners that enable the exposure of the ultraviolet (UV) image of a mask pattern onto a substrate coated with photoresist (PR). In general, the equipment used in academic institutes is the contact printing type for relatively low-resolution applications. Although the resolution of the contact aligner is not very high, it is expensive and bulky, because it employs a high-power mercury lamp with a bulky cooling system. In addition, the use of the contact mask aligner requires the use of a fine-patterned photo-mask. Since the pattern on the mask is directly transferred onto the substrate, the mask must have the same resolution or precision as the target pattern. Since a high-resolution mask requires e-beam lithography or the laser directly writing on the chrome mask, it is significantly time-consuming and expensive to acquire.

In this paper, we will deal with microscope projection photolithography (MPP) and the results of its implementation with a UV light-emitting diode (LED) as a light source. MPP has a long history of the development and applications in the research area [2–5]. Traditionally, MPP utilizes a mercury lamp as a light source for the contact printer. It may cause the same problems as the contact aligner in terms of the cost and size of the equipment.

The use of a UV LED as the light source for MPP can bring several advantages in terms of cost and size. These days, high-power UV LEDs are readily available and very cheap. Since the size of a UV LED is very small, the projection instrument can match the compactness of the microscope. Furthermore, the MPP has the capability of reducing the image using a magnified objective lens. This means that the MPP can use a low-resolution photomask, which can be easily made with an image setter, a low-resolution laser writer, or even an ink jet printer. The alleviation of the burden of a high-resolution photomask enables rapid prototyping in the experiment. In addition, since MPP is a type of projection printing, it also inherits the advantages of non-contact printing, such as the contamination-free exposure.

In this study, we are going to investigate the feasibility of UV LED as a light-source for MPP and check the performance to an extent by making the MPP equipment and finding the best photolithography conditions. In this manner, we would like to demonstrate the suitability of the MPP based on UV LEDs as equipment for education and research.

## 2. Description of the method

The experiment consists of the adaptation of a microscope for projection photolithography and the establishment of the photolithography conditions. The photolithography process include the coating of the PR on the glass substrate, fabrication of the photomasks, UV exposure, development of the PR patterns, chemical etching, and removal of the PR. The repetition of the photolithography process leads to the optimization of the process through the verification of the size limit of the PR patterns.

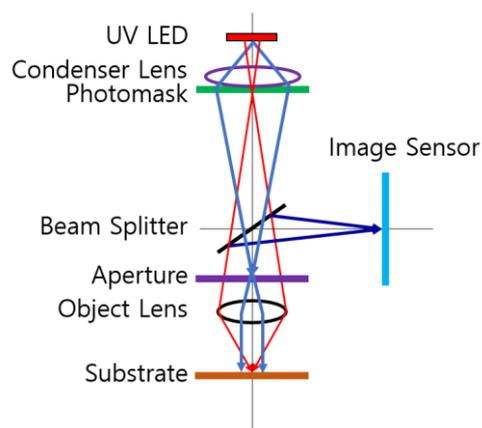

Fig. 1. Schematic of the adaptation for the projection photolithography.

The adaptation of a microscope into a projection system demands the knowledge of Köhler illumination for the microscope[6]. The schematic diagram shown in Fig. 1 may help us to understand the mechanism of the microscope adapted as a projector using a UV LED. In this configuration, the photomask, the substrate, and the image sensor are all conjugate to each other. The UV LED is imaged on the aperture stop of the objective lens, which is usually the back focal plane, contributing to the uniform illumination over the substrate.

A mechanical structure holding the UV LED, condenser lens, and photomask was machined and assembled, as shown in Fig. 2(a). Fig. 2(b), (c), and (d) show the condenser lens, the photomask, and the UV LED installed in their respective holders. Fig. 2(e) shows the UV image of the grating pattern on the substrate position when using a 2X objective lens. The image sensor located in the reflected optical path can be used to confirm the alignment between the sample and the image of the photomask.

As for the UV LED used in the adapted microscope, the exact specifications were not well known since it was purchased at a very low price from an online marketplace. The maximum electrical power was about 1 W and the wavelength was around 390 nm.

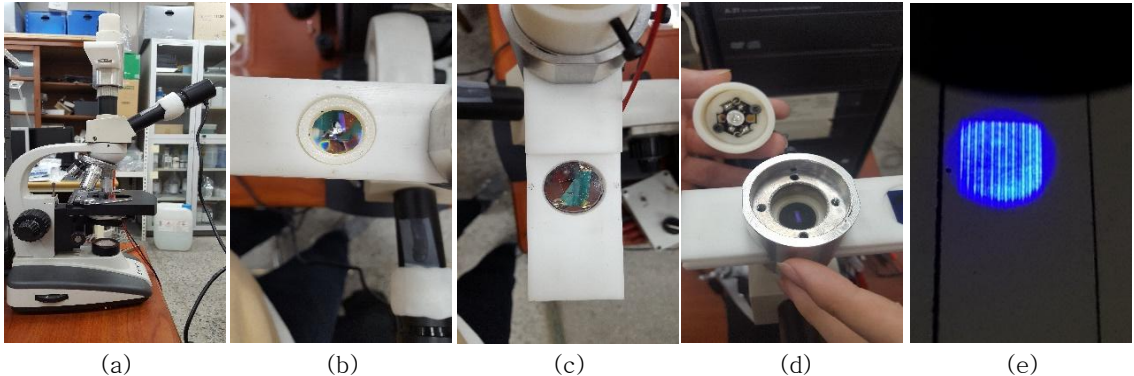

(a)  (b)  (c)  (d)  (e)

Fig. 2. Parts of the optics module and the UV image of a photomask: (a) the adapted microscope for the projection photolithography, (b) the condenser lens for collecting the light from a UV LED, (c) the photomask installed near condenser lens, (d) the UV LED as the light source, and (e) an image of the photomask at the position of the substrate. A paper was used in place of the substrate for visualization.

The photolithography using the adapted microscope started with the coating of silver onto a glass plate. The sputtering machine evaporated silver onto a glass plate, with the thickness estimated to be about 200 nm. The silver-coated glass plate went through spin coating with the PR at 3500 rpm for 30 s. The soft bake of the PR was carried out on a hot plate at 100 ℃ for 180 s. The PR used in the experiment was obtained by thinning AZ4562 with PGMEA (2-methoxy-1-methylethylacetate).

The second step of the experiment was the fabrication of the photomask using the doctor blade method[7-8] and an ink jet. The doctor blade method is also called the tape casting method because the glass plate has two tapes on it, and the application of silver paint between the two tapes is followed by scraping off the excess paint with a blade for the formation of a uniform film. After the silver paint dries up, the cutting blade is used for making a narrow slit on the sliver coating. The narrow slit acts as a photomask to check the lithography resolution and the processes.

A second method for making a photomask is to use an ink-jet printer. A conventional ink-jet printer can print the images on a transparency film, as shown in Fig. 3. The design of the patterns can be easily done by common software, which is also a useful feature for rapid prototyping.

In order to obtain high contrast patterns from the MPP, we focused on the optimization of the UV exposure time and developing time. At first, the 2X objective lens was used to image the photomask on the substrate coated with PR. The magnification factor in the MPP represents the reduction ratio of the image on the substrate to the pattern of the photomask. It also affects the irradiance of the UV light on the substrate. Theoretically, the irradiance is proportional to the square of the magnification factor.

After UV exposure for a given time, the sample was developed with the AZ400 developer for a certain time. The time of development and UV exposure were varied to achieve the best resolution of PR pattern.

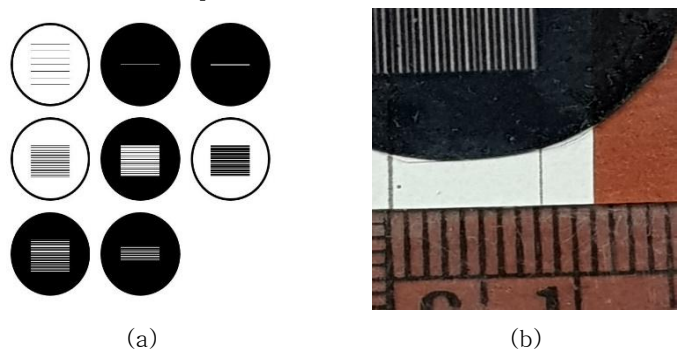

(a) (b)

Fig. 3. (a) Photomask patterns designed on a computer. (b) Photomask patterns printed by an ink-jet printer, with a ruler where the smallest unit is 1 mm.

The etching of the silver coating follows the development process. The etching solution for the silver was prepared by mixing 100 mL of water, 4.15 mL of ammonium hydroxide, and 4.13 mL of hydrogen peroxide[9]. The sample with the PR pattern was immersed in the etching solution and stirred slowly to help the diffusion of the etching chemicals onto the sample surface. After the etching, the sample was rinsed with water, dried using a nitrogen gun, and underwent inspection under a microscope for the evaluation of the pattern quality. The optimization of the lithography conditions was carried out by repeating the procedures with variations in exposure time, development time, and etching time until a good quality pattern was obtained. After the good quality sample was obtained, the PR on the substrate was removed by spraying it with acetone.

Each sample has a picture taken by the microscope's digital camera for the evaluation of the lithography conditions. The picture was calibrated by the photo of a microscope ruler with 10-micron markings per division under the same magnification of the objective lens and video frame as the measured sample. Since the conversion rate from pixels to microns is different in the horizontal and vertical directions, the calibration and measurement was mainly performed in the horizontal direction.

The quality of the lithography was chiefly assessed by checking the line widths and the quality of the slit pattern after development, etching, and PR removal.

In order to investigate the resolution limit of the adapted microscope projector based on UV LEDs, we tried to use a 4X objective lens as well as a 2X objective lens. A higher magnification objective lens generally results in a higher reduction ratio and higher irradiance of UV light onto the sample. Thus, the optimization of the photolithography conditions was carried out separately for each case.

Another variation in the lithography conditions of the experiment was made with the thickness of the PR. The thickness of the PR can be controlled by mixing the thinner in different ratios. The thick PR and the thin PR were named A and C, respectively. The PR named B was made by mixing A and C at a ratio of 3 to 1. Unfortunately, the exact mix ratios between the thinner and AZ4562 were not recorded when PR A and PR B were prepared.

## 3. Results and Discussion

The first experiment focused on a single slit pattern. The photomask of a single slit was made by the doctor blade method mentioned before. The six samples experienced different UV exposure times, developing times, and etching times, as shown in Table 1. For the samples with exposure times of 30 s and developing times of 60 s, the glass surfaces were not completely exposed after etching, regardless of the etching time. The same thing happens for samples with exposure times of 45 s or developing times of 90 s. As for the samples with exposure times greater than 60 s and developing times greater than 120 s, the glass is exposed as a clean surface as the etching time increases. The clean surface generally indicates the completion of the silver etching. When the etching time reaches 50 s or 60 s, the surface becomes almost clear without sliver residue. Therefore, the samples fabricated by the silver paint mask have optimal lithography conditions with an exposure time of 60 s, a developing time of 120 s, and an etching time of 60 s. The microscopic images of the four samples after the fabrication were seen in Fig. 4.

Table 1. Photolithography conditions used for the single slit patterns. The conditions differ in terms of the UV exposure, developing time, and etching time. The photomask has two separated lines on the sliver-painted glass plate.

| Sample | UV exposure (s) | Developing time (s) | Etching time (s) | Sample | UV exposure (s) | Developing time (s) | Etching time (s) |
|---|---|---|---|---|---|---|---|
| A-1 | 30 | 60 | 30 | D-1 | 45 | 90 | 30 |
| A-2 | 30 | 60 | 50 | D-2 | 45 | 90 | 50 |
| A-3 | 30 | 60 | 60 | D-3 | 45 | 90 | 60 |
| B-1 | 30 | 90 | 30 | E-1 | 60 | 120 | 30 |
| B-2 | 30 | 90 | 50 | E-2 | 60 | 120 | 50 |
| B-3 | 30 | 90 | 60 | E-3 | 60 | 120 | 60 |
| C-1 | 45 | 60 | 30 | F-1 | 120 | 120 | 30 |
| C-2 | 45 | 60 | 50 | F-2 | 120 | 120 | 50 |
| C-3 | 45 | 60 | 60 | F-3 | 120 | 120 | 60 |

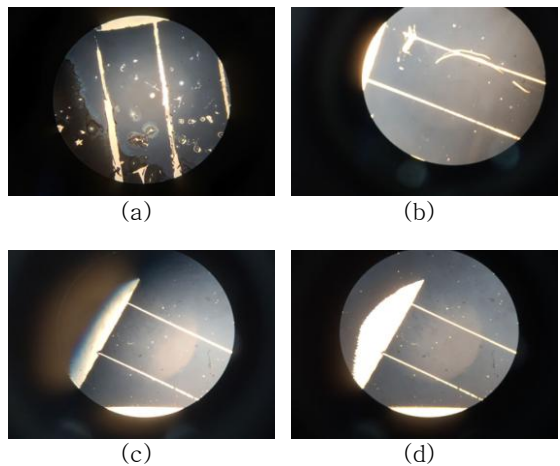

Fig. 4. Microscopic images of the two single slit patterns after PR removal: (a) A-1, (b) C-1, (c) E-3, and (d) F-3.

However, the best lithography conditions for the single slit pattern were not suitable for the grating patterns made by the ink-jet-printed photomask. An increase in the developing time to 180 s did not greatly improve the quality of patterns. As the exposure time increased to 120 s and the developing time increased to 180 s and then 240 s, the pattern improved a little, with small PR residues on the surface. When the exposure time increased up to 150 s, the substrate exposed the clean surface of the glass. The etching condition did not depend on the type of photomask. The etching was carried out for 60 s as in the first experiment with a sliver painted mask. The lithography conditions for the grating patterns used in the experiment are summarized in Table 2. The microscopic images after fabrication are shown in Fig. 5.

Table 2. Photolithography conditions used for the grating patterns. The conditions differ in terms of the UV exposure, developing time, and etching time. The photomask has the grating pattern on a transparency film printed by an inkjet.

| Sample | UV exposure (s) | Developing time (s) | Etching time (s) |
|---|---|---|---|
| G-1 | 90 | 120 | 60 |
| G-2 | 90 | 180 | 60 |
| H-1 | 120 | 180 | 60 |
| H-2 | 120 | 240 | 60 |
| H-3 | 150 | 180 | 60 |

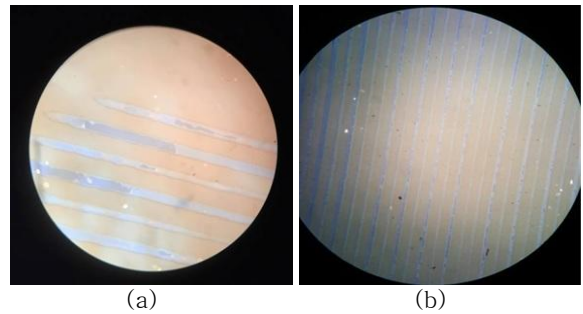

Fig. 5. Microscopic images of the grating patterns after PR removal: (a) G-1 and (b) H-2.

The measurements of the fabricated samples indicate that the final pattern agrees well with the expected reduction ratio estimated from the magnification factor of the microscope objective lens. The final pattern width of the single slit was 17 μm, which was about half that of the one in the photomask shown in Fig. 6. The same reduction occurred for the grating pattern, as seen in Fig. 7.

In order to find out how a small pattern can be made by the adapted microscope projector, a smaller mask pattern was used and the corresponding optimization was carried out. As mentioned before, the single slit pattern with a width of 17 μm fabricated in the previous experiment was used as the photomask, as shown in Fig. 8. The use of 2X and 4X objective lenses yielded the reduction ratios in the final pattern. Since changing the objective lens affects the lithography conditions, this requires a new search for the best conditions for the process. The parameters used for the process optimization are summarized in Table 3.

| microscope ruler | photomask |
|---|---|
| 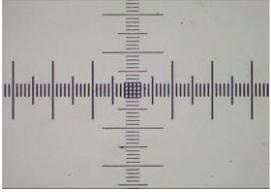 | 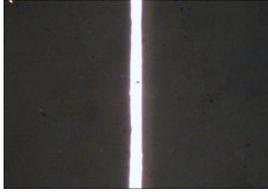 |
| (a) | (b) |
| 96 pixels = 100μm | 29 pixels = 30 μm |
| Sample | |
| 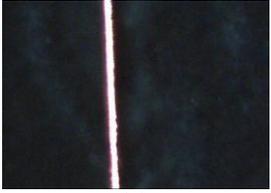 | |
| (c) | |
| 16 pixels = 17 μm | |

Fig. 6. Single slit pattern fabricated by the MPP.
(a) Microscope ruler used for calibration, (b) the photomask, and (c) the fabricated sample.

| microscope ruler | photomask |
|---|---|
| 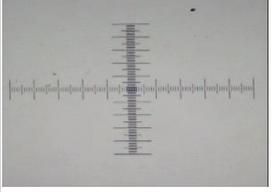 | 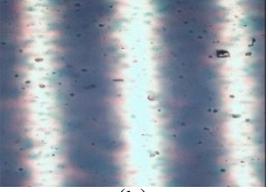 |
| (a) | (b) |
| 42 pixels = 100μm | 70 pixels = 170 μm |
| Sample | |
| 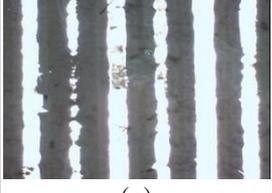 | |
| (c) | |
| 35 pixels = 83 μm | |

Fig. 7. Grating pattern fabricated by the MPP. (a) Microscope ruler used for calibration, (b) the photomask, and (c) the fabricated sample.

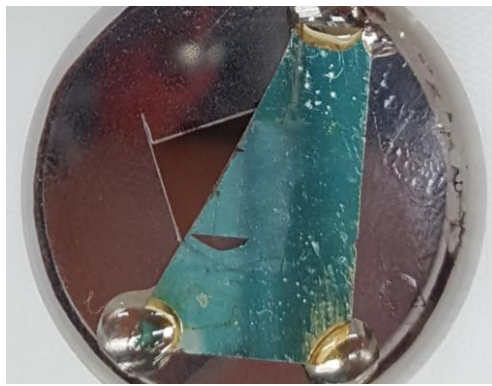

Fig. 8. 17-μm-wide single slit pattern fixed at the aperture in order to be used as a photomask. This photomask was used to test a higher resolution pattern.

Table 3. Photolithography conditions used for checking the higher resolution limit.

| Sample | PR | Soft bake (s) | UV exposure (s) | Developing time (s) | Etching time (s) | Magnification |
|---|---|---|---|---|---|---|
| 1 | A | 180 | 90 | 120 | 60 | 2 |
| 2 | C | 180 | 60 | 60 | 60 | 2 |
| 3 | A | 180 | 150 | 120 | 60 | 2 |
| 4 | C | 180 | 90 | 60 | 60 | 2 |
| 5 | A | 60 | 90 | 150 | 60 | 2 |
| 6 | B | 60 | 90 | 90 | 20 | 2 |
| 7 | B | 60 | 30 | 90 | 20 | 4 |
| 8 | A | 60 | 60 | 90 | 30 | 4 |

The microscopic images of the samples 6 and 8 are shown in Fig. 9. Although the line width of sample 8 is about half that of sample 6 after development, the line width of sample 8 becomes similar to that of sample 6 after PR removal. This can also be confirmed in Table 4 according to the measured quantitative data. The widening of the line width after PR removal can be ascribed to the undercut etch. The etchant usually penetrates to between the PR mask and the metal layer on the substrate. Therefore, the resolution limit of the MPP can be correctly assessed by the line width of the PR pattern right after development.

The measurements of the line widths for each sample are summarized in Table 4. According to this table, the smallest pattern was obtained from sample 8. The line width was 5.0 μm after development. This sample was projected by a 4X objective lens and the line width is about half of that of sample 6, which was obtained from a 2X objective lens. The experimental results show that the reduction rule remains valid as the magnification factor of microscope objective reaches as high as 4. The diffraction-limited resolution is the wavelength times the f-number of the lens, according to the Rayleigh criterion[10]. The f-number of the 4X objective lens is 5.0 since its numerical aperture was 0.10. The theoretical resolution is 2.0 μm, assuming that the wavelength is 0.40 μm. Since this resolution limit represents the minimum separation of two points with zero contrast, the actual resolution limit is about 4 μm, which is close to the minimum line width measured in our experiment.

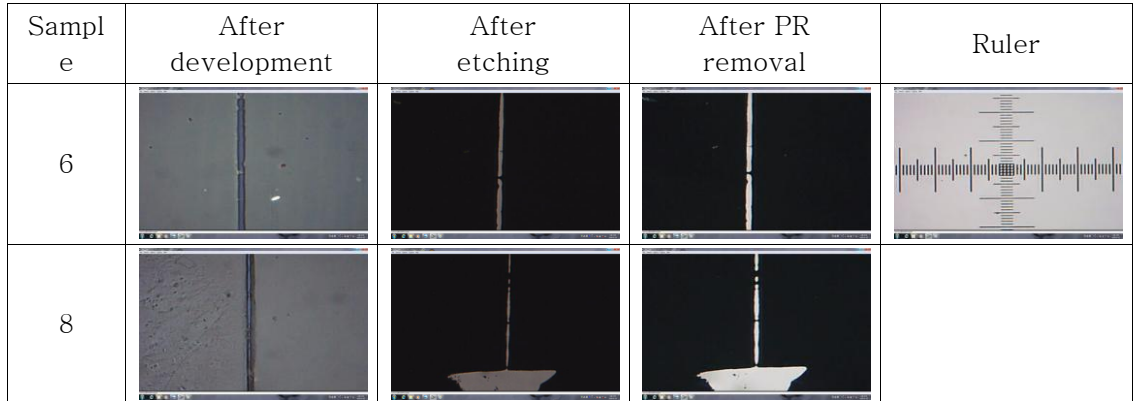

Fig. 9. Microscopic images of the line patterns after development, etching, and PR removal. The sample numbers are from Table 3.

Table 4. Line widths measured after the development, etching, and PR removal. One pixel in the original photo corresponds to 0.334 μm.

| Sample | After Development (pixel) | After Development (μm) | After etching (pixel) | After etching (μm) | After PR removal (pixel) | After PR removal (μm) |
|---|---|---|---|---|---|---|
| 1 | 24 | 8.0 | 43 | 14 | 47 | 16 |
| 2 | 29 | 9.7 | 39 | 13 | 44 | 15 |
| 3 | 27 | 9.0 | 43 | 14 | 44 | 15 |
| 4 | 25 | 8.4 | 37 | 12 | 41 | 14 |
| 5 | 50 | 17 | 51 | 17 | 69 | 23 |
| 6 | 32 | 11 | 32 | 11 | 34 | 11 |
| 7 | 19 | 6.4 | 21 | 7.0 | 28 | 9.4 |
| 8 | 15 | 5.0 | 16 | 5.4 | 27 | 9.0 |

Finally, we also applied the fabricated single slit to the diffraction experiment, as shown in Fig. 10. The observed diffraction pattern agrees well with the expected profile from the measured slit width. It shows an example of how to apply MPP to physics education. The microscope adapted for the projection photolithography can make a pattern as small as the diffraction limit allows. According to the literature[2], the feature size can be reduced to 1 micron or even less. Therefore, the low-cost MPP based on UV LEDs can contribute to the field of physics education or various areas of research, such as chemistry and biology.

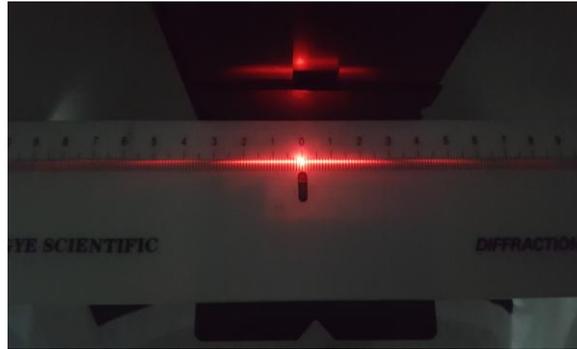
Fig. 10. Diffraction experiment was carried out using the fabricated single slit.

## 4. Conclusions

We adapted a conventional microscope for projection photolithography using a UV LED as the light source. The use of a UV LED provides the microscope projector with many advantages in terms of compactness and cost. The adaptation of the microscope followed the principles of Köhler illumination and the LED replaced the mercury lamp in the older version of the MPP. The UV LED, a condenser lens, and a photomask were integrated into a mechanical module. This optical module enabled the microscope to project the photomask onto a substrate coated with PR.

We tested the performance of the adapted microscope as a lithography projector. The testing required the optimization of the photolithography conditions as well as the imaging capabilities of the microscope. The photomasks used in the experiment were single slit patterns on silver-painted glass and a grating pattern on a transparency printed by an ink-jet printer. Both patterns were made successfully, showing line widths of 17 μm and 83 μm, respectively. These results also demonstrated a two-fold reduction when a 2X objective lens was used.

In order to check whether the pattern size could be reduced further, a 4X objective lens and different PR thicknesses were tried with the corresponding changes in the lithography processes. The smallest line width of the final PR pattern was 5.0 μm when the photomask of a 17-μm-wide line was applied. It turns out that the final line width was close to the diffraction limit of the lens.

Therefore, the use of a finer photomask and higher magnification of the objective lens may lead to a smaller line width. This is a surprising result, considering that the adapted microscope employed a very cheap and compact UV LED and a very crude mechanical structure for the optical module. We expect that the low-cost MPP based on UV LED can make contributions in the field of physics education or various areas of research, such as chemistry and biology, in the future.

### Acknowledgements
We appreciate the help of Haneol Park, Youngjin Jung and Kyungjun Oh and their involvement in the beginning of this experiment.